\documentclass[aps,prb,twocolumn,superscriptaddress,nofootinbib,floatfix]{revtex4-2}

\usepackage{amsmath,amssymb,bm}
\usepackage{booktabs}
\usepackage{graphicx}
\usepackage{tikz}
\usetikzlibrary{arrows.meta,calc,decorations.markings,decorations.pathmorphing,positioning,shapes.geometric}
\usepackage[colorlinks=true,linkcolor=blue,citecolor=blue,urlcolor=blue]{hyperref}

\definecolor{qred}{HTML}{D55E00}
\definecolor{qblue}{HTML}{0072B2}
\definecolor{qgreen}{HTML}{009E73}
\definecolor{qorange}{HTML}{E69F00}
\definecolor{qgray}{HTML}{5B6573}
\tikzset{
  qarrow/.style={-{Latex[length=2.0mm,width=1.35mm]},line width=0.8pt},
  flow/.style={-{Latex[length=2.0mm,width=1.35mm]},line width=0.9pt},
  domainbox/.style={draw=qgray,rounded corners=2pt,fill=qgray!5,
    minimum width=2.15cm,minimum height=0.72cm,align=center},
  groupbox/.style={draw=qgray,rounded corners=3pt,fill=white,
    minimum width=2.1cm,minimum height=0.8cm,inner xsep=4pt,align=center},
  note/.style={font=\scriptsize,align=center,text=qgray}
}

\newcommand{\QQ}{\bm{Q}}
\newcommand{\MM}{\bm{M}}
\newcommand{\ZZ}{\mathbb{Z}}
\newcommand{\OO}{\mathrm{O}}
\newcommand{\SO}{\mathrm{SO}}
\newcommand{\SU}{\mathrm{SU}}
\newcommand{\Dthree}{D_3}
\newcommand{\Vfour}{V_4}

\begin{document}

\title{Topological Defects in Triple-$Q$ Magnetic Orders: A Fixed-Lattice Homotopy Classification}

\author{Jin-Tao Jin}
\affiliation{Department of Physics, The Hong Kong University of Science and Technology, Clear Water Bay, Kowloon 999077, Hong Kong, China}

\author{Yi Zhou}
\email{yizhou@iphy.ac.cn}
\affiliation{Institute of Physics, Chinese Academy of Sciences, Beijing 100190, China}

\begin{abstract}
Multiple-$Q$ magnetic orders combine continuous spin rotations with discrete crystalline sectors associated with translations and point-group transformations, producing a richer defect structure than conventional single-$Q$ magnets. We classify the bulk defects of all seven stable phases for $N=2$ and $3$ in the $M$-point triple-$Q$ Ginzburg--Landau theory with $(\Vfour\rtimes\Dthree)\times\OO(N)$ symmetry, where $\Vfour$ is the translation-generated Klein four-group. The atomic lattice is treated as a prescribed background, with lattice dislocations and disclinations excluded and the three Fourier fields retaining their physical $M$-point labels. The parent-group transformations continuously connected to the identity form $G_0=\{e\}\times\SO(N)$. For a reference-state stabilizer $H$, the connected component containing the reference state is $G_0/(H\cap G_0)$, not the quotient obtained by projecting $H$ onto spin space. This distinction gives the orthogonal triple-$Q$ phase the full manifold $\OO(3)$, with chirality walls and Abelian $\ZZ_2$ frame vortices rather than non-Abelian binary-polyhedral vortices. Every connected component of the $\OO(2)$ phases supports an integer $2\pi$ vortex, whereas fractional windings close only when attached to a discrete-domain wall and are linearly confined at nonzero wall tension. Translation symmetry further forbids cross-gradient bilinears, reducing the quadratic elastic sector to an isotropic and an $M$-point-locked anisotropic stiffness. The classification separates free internal defects, crystalline domain walls, and wall-bound composites in triple-$Q$ magnets.
\end{abstract}

\maketitle

\section{Introduction}
\label{sec:introduction}

Multiple-$Q$ magnetic order arises naturally from frustrated exchange and itinerant magnetic interactions~\cite{Batista2016,HayamiMotome2021,KatoPRB2022,Messio2011}. Noncoplanar triple-$Q$ states are particularly notable because their scalar spin chirality can generate spontaneous Hall responses~\cite{MartinBatista2008,Kato2010PRL,Kumar2010}. Triple-$Q$ order has been proposed and experimentally reported in the honeycomb cobaltate Na$_2$Co$_2$TeO$_6$~\cite{YaoPRR2023,Kruger2023,Francini2024PRB,JinExperiment2025}, while a tetrahedral triple-$Q$ state has been identified in the metallic triangular antiferromagnet Co$_{1/3}$TaS$_2$~\cite{Park2023}. For ordering at the three symmetry-related $M$ points of a hexagonal Brillouin zone, the physically labeled Fourier amplitudes may form collinear, coplanar $120^\circ$, or mutually orthogonal spin configurations. Such $M$-point order enlarges the magnetic unit cell and admits translation-related antiphase domains. The ordered-state space therefore combines continuous spin degrees of freedom with discrete crystalline sectors, making the closure conditions for topological defects nontrivial.

Topological defects are classified by the homotopy groups of the ordered-state manifold~\cite{ToulouseKleman1976,Mermin1979,Michel1980,KawaguchiUeda2012}. For a reference ordered state $\Phi_0$, this manifold is the symmetry orbit $\mathcal M=G/H$, where the stabilizer $H\subset G$ consists of all parent-group operations that leave $\Phi_0$ unchanged. The connected components of $\mathcal M$, denoted $\pi_0(\mathcal M)$, label distinct domain sectors. Within a chosen component $\mathcal M_0$, $\pi_1(\mathcal M_0)$ classifies noncontractible loops around vortex cores, whereas $\pi_2(\mathcal M_0)$ classifies skyrmion textures in two dimensions and hedgehog defects in three. A complication arises when $G$ is disconnected: $H$ may contain a combined operation $(s,R)$, where $s$ is a translation or point-group operation and $R$ is a spin rotation, even though $R$ alone does not fix $\Phi_0$. Instead, $R\Phi_0$ is the distinct state obtained by acting with $s^{-1}$. Projecting $H$ onto spin space would incorrectly treat $R$ as a stabilizer and make an open path in $\mathcal M$ appear closed, producing spurious fractional or non-Abelian defect charges.

We present a fixed-lattice classification of the seven stable phases of the triple-$Q$ Ginzburg--Landau theory introduced in Ref.~\cite{Jin2025}. The atomic lattice is a prescribed, nondynamical background, and the three order parameters are physically labeled Fourier amplitudes at fixed $M$-point momenta. Translations and point-group operations therefore act as global symmetries rather than gauge identifications. A free internal-defect loop must return to the same physical order-parameter state with every $M$-point label restored. Closure after a discrete crystalline transformation requires a domain wall, whereas closure through a lattice dislocation or disclination lies outside the fixed-lattice classification. For an $N$-component order parameter, the component containing $\Phi_0$ is generated by $G_0=\{e\}\times\SO(N)$, where $e$ is the identity crystalline operation. Its stabilizer is $K=H\cap G_0$, not the spin-space projection of $H$.

This construction changes the resulting defect classification. The orthogonal $N=3$ triple-$Q$ phase IIIB supports chirality walls and Abelian $\ZZ_2$ frame vortices, rather than non-Abelian binary-polyhedral vortices familiar from multicomponent ordered media~\cite{Kobayashi2009}. Every connected component of each $N=2$ phase supports an integer $2\pi$ vortex. A $\pi$ winding can become single valued only when attached to a discrete-domain wall and is therefore not an element of the bulk fundamental group. We also derive the quadratic elastic theory, in which translation symmetry forbids cross-label bilinears, leaving an isotropic stiffness and an $M$-point-locked anisotropic stiffness.

The remainder of the paper is organized as follows. Section~\ref{sec:model} introduces the Ginzburg--Landau model, symmetry action, and fixed-lattice convention. Section~\ref{sec:homotopy} develops the homotopy construction for a disconnected parent group, and Sec.~\ref{sec:classification} applies it to all seven stable phases. Section~\ref{sec:elasticity} derives the symmetry-allowed elastic terms. Section~\ref{sec:discussion} discusses wall-bound configurations, observable consequences, and extensions beyond the ideal fixed-lattice setting, and Sec.~\ref{sec:conclusion} summarizes the results.

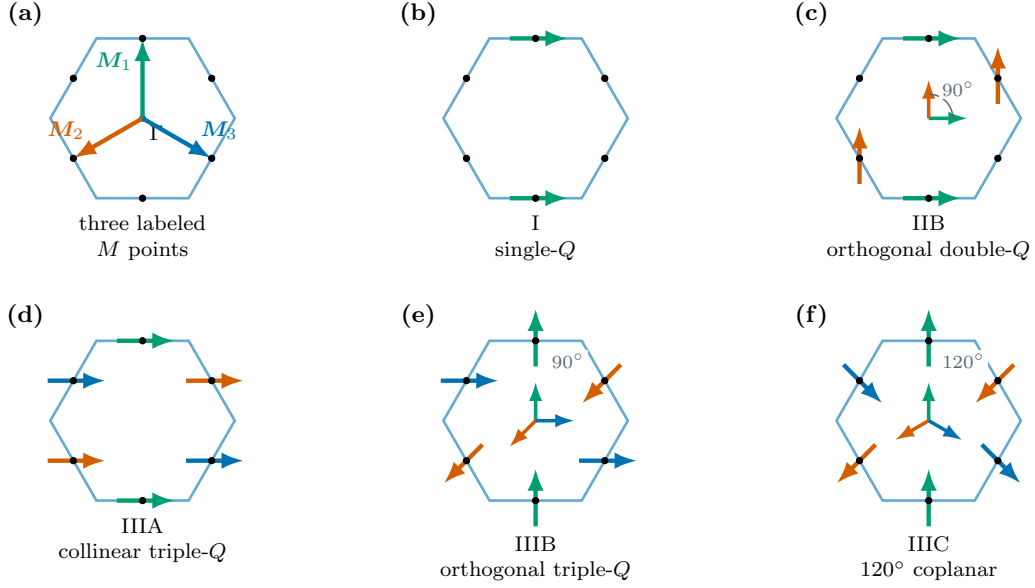
\begin{figure*}[tb]
\centering
\begin{tikzpicture}[
  x=1cm,y=1cm,font=\footnotesize,
  bz/.style={draw=qblue!60,line width=1.0pt},
  spin/.style={-{Latex[length=3.0mm,width=2.2mm]},line width=1.65pt},
  triad/.style={-{Latex[length=2.6mm,width=1.8mm]},line width=1.35pt},
  phase/.style={font=\footnotesize,align=center},
  angle/.style={font=\scriptsize,text=qgray,fill=white,inner sep=1.2pt}
]

  \begin{scope}[shift={(0,2.05)}]
    \node[font=\bfseries] at (-1.55,1.38) {(a)};
    \draw[bz]
      (0:1.22) \foreach \a in {60,120,...,300} {-- (\a:1.22)} -- cycle;
    \foreach \p in {(0,1.057),(0,-1.057),(0.915,0.529),(0.915,-0.529),
      (-0.915,0.529),(-0.915,-0.529)}{
      \fill \p circle (1.25pt);
    }
    \coordinate (G) at (0,0);
    \coordinate (M1) at (0,1.057);
    \coordinate (M2) at (-0.915,-0.529);
    \coordinate (M3) at (0.915,-0.529);
    \fill (G) circle (1.25pt);
    \node[below right=-1pt] at (G) {$\Gamma$};
    \draw[spin,qgreen] (G) -- (M1)
      node[pos=0.72,left] {$\MM_1$};
    \draw[spin,qred] (G) -- (M2)
      node[pos=0.72,above left=-1pt] {$\MM_2$};
    \draw[spin,qblue] (G) -- (M3)
      node[pos=0.72,above right=-1pt] {$\MM_3$};
    \foreach \p in {(0,1.057),(0,-1.057),(0.915,0.529),(0.915,-0.529),
      (-0.915,0.529),(-0.915,-0.529)}{
      \fill \p circle (1.35pt);
    }
    \node[phase] at (0,-1.58) {three labeled\\$M$ points};
  \end{scope}

  \begin{scope}[shift={(5.2,2.05)}]
    \node[font=\bfseries] at (-1.55,1.38) {(b)};
    \draw[bz]
      (0:1.22) \foreach \a in {60,120,...,300} {-- (\a:1.22)} -- cycle;
    \coordinate (t) at (0,1.057);
    \coordinate (b) at (0,-1.057);
    \foreach \p in {(0,1.057),(0,-1.057),(0.915,0.529),(0.915,-0.529),
      (-0.915,0.529),(-0.915,-0.529)}{
      \fill \p circle (1.25pt);
    }
    \foreach \p in {t,b}{
      \draw[spin,qgreen] ($(\p)+(-0.34,0)$) -- ($(\p)+(0.43,0)$);
    }
    \foreach \p in {(0,1.057),(0,-1.057),(0.915,0.529),(0.915,-0.529),
      (-0.915,0.529),(-0.915,-0.529)}{
      \fill \p circle (1.35pt);
    }
    \node[phase] at (0,-1.58) {I\\single-$Q$};
  \end{scope}

  \begin{scope}[shift={(10.4,2.05)}]
    \node[font=\bfseries] at (-1.55,1.38) {(c)};
    \draw[bz]
      (0:1.22) \foreach \a in {60,120,...,300} {-- (\a:1.22)} -- cycle;
    \coordinate (t) at (0,1.057);
    \coordinate (b) at (0,-1.057);
    \coordinate (ur) at (0.915,0.529);
    \coordinate (ll) at (-0.915,-0.529);
    \foreach \p in {(0,1.057),(0,-1.057),(0.915,0.529),(0.915,-0.529),
      (-0.915,0.529),(-0.915,-0.529)}{
      \fill \p circle (1.25pt);
    }
    \foreach \p in {t,b}{
      \draw[spin,qgreen] ($(\p)+(-0.34,0)$) -- ($(\p)+(0.43,0)$);
    }
    \foreach \p in {ur,ll}{
      \draw[spin,qred] ($(\p)+(270:0.34)$) -- ($(\p)+(90:0.43)$);
    }
    \draw[triad,qgreen] (0,0) -- (0:0.52);
    \draw[triad,qred] (0,0) -- (90:0.52);
    \draw[qgray,line width=0.55pt] (0.32,0) arc[start angle=0,end angle=90,radius=0.32];
    \node[angle] at (0.39,0.39) {$90^\circ$};
    \foreach \p in {(0,1.057),(0,-1.057),(0.915,0.529),(0.915,-0.529),
      (-0.915,0.529),(-0.915,-0.529)}{
      \fill \p circle (1.35pt);
    }
    \node[phase] at (0,-1.58) {IIB\\orthogonal double-$Q$};
  \end{scope}

  \begin{scope}[shift={(0,-1.95)}]
    \node[font=\bfseries] at (-1.55,1.38) {(d)};
    \draw[bz]
      (0:1.22) \foreach \a in {60,120,...,300} {-- (\a:1.22)} -- cycle;
    \coordinate (t) at (0,1.057);
    \coordinate (b) at (0,-1.057);
    \coordinate (ur) at (0.915,0.529);
    \coordinate (ll) at (-0.915,-0.529);
    \coordinate (ul) at (-0.915,0.529);
    \coordinate (lr) at (0.915,-0.529);
    \foreach \p in {(0,1.057),(0,-1.057),(0.915,0.529),(0.915,-0.529),
      (-0.915,0.529),(-0.915,-0.529)}{
      \fill \p circle (1.25pt);
    }
    \foreach \p in {t,b}{
      \draw[spin,qgreen] ($(\p)+(-0.34,0)$) -- ($(\p)+(0.43,0)$);
    }
    \foreach \p in {ur,ll}{
      \draw[spin,qred] ($(\p)+(-0.34,0)$) -- ($(\p)+(0.43,0)$);
    }
    \foreach \p in {ul,lr}{
      \draw[spin,qblue] ($(\p)+(-0.34,0)$) -- ($(\p)+(0.43,0)$);
    }
    \foreach \p in {(0,1.057),(0,-1.057),(0.915,0.529),(0.915,-0.529),
      (-0.915,0.529),(-0.915,-0.529)}{
      \fill \p circle (1.35pt);
    }
    \node[phase] at (0,-1.58) {IIIA\\collinear triple-$Q$};
  \end{scope}

  \begin{scope}[shift={(5.2,-1.95)}]
    \node[font=\bfseries] at (-1.55,1.38) {(e)};
    \draw[bz]
      (0:1.22) \foreach \a in {60,120,...,300} {-- (\a:1.22)} -- cycle;
    \coordinate (t) at (0,1.057);
    \coordinate (b) at (0,-1.057);
    \coordinate (ur) at (0.915,0.529);
    \coordinate (ll) at (-0.915,-0.529);
    \coordinate (ul) at (-0.915,0.529);
    \coordinate (lr) at (0.915,-0.529);
    \foreach \p in {(0,1.057),(0,-1.057),(0.915,0.529),(0.915,-0.529),
      (-0.915,0.529),(-0.915,-0.529)}{
      \fill \p circle (1.25pt);
    }
    \foreach \p in {t,b}{
      \draw[spin,qgreen] ($(\p)+(270:0.34)$) -- ($(\p)+(90:0.43)$);
    }
    \foreach \p in {ur,ll}{
      \draw[spin,qred] ($(\p)+(45:0.30)$) -- ($(\p)+(225:0.43)$);
    }
    \foreach \p in {ul,lr}{
      \draw[spin,qblue] ($(\p)+(180:0.34)$) -- ($(\p)+(0:0.43)$);
    }
    \draw[triad,qgreen] (0,0) -- (90:0.52);
    \draw[triad,qred] (0,0) -- (225:0.50);
    \draw[triad,qblue] (0,0) -- (0:0.52);
    \node[angle] at (0.43,0.79) {$90^\circ$};
    \foreach \p in {(0,1.057),(0,-1.057),(0.915,0.529),(0.915,-0.529),
      (-0.915,0.529),(-0.915,-0.529)}{
      \fill \p circle (1.35pt);
    }
    \node[phase] at (0,-1.82) {IIIB\\orthogonal triple-$Q$};
  \end{scope}

  \begin{scope}[shift={(10.4,-1.95)}]
    \node[font=\bfseries] at (-1.55,1.38) {(f)};
    \draw[bz]
      (0:1.22) \foreach \a in {60,120,...,300} {-- (\a:1.22)} -- cycle;
    \coordinate (t) at (0,1.057);
    \coordinate (b) at (0,-1.057);
    \coordinate (ur) at (0.915,0.529);
    \coordinate (ll) at (-0.915,-0.529);
    \coordinate (ul) at (-0.915,0.529);
    \coordinate (lr) at (0.915,-0.529);
    \foreach \p in {(0,1.057),(0,-1.057),(0.915,0.529),(0.915,-0.529),
      (-0.915,0.529),(-0.915,-0.529)}{
      \fill \p circle (1.25pt);
    }
    \foreach \p in {t,b}{
      \draw[spin,qgreen] ($(\p)+(270:0.34)$) -- ($(\p)+(90:0.43)$);
    }
    \foreach \p in {ur,ll}{
      \draw[spin,qred] ($(\p)+(45:0.30)$) -- ($(\p)+(225:0.43)$);
    }
    \foreach \p in {ul,lr}{
      \draw[spin,qblue] ($(\p)+(135:0.30)$) -- ($(\p)+(315:0.43)$);
    }
    \draw[triad,qgreen] (0,0) -- (90:0.52);
    \draw[triad,qred] (0,0) -- (210:0.52);
    \draw[triad,qblue] (0,0) -- (330:0.52);
    \node[angle] at (0.46,0.79) {$120^\circ$};
    \foreach \p in {(0,1.057),(0,-1.057),(0.915,0.529),(0.915,-0.529),
      (-0.915,0.529),(-0.915,-0.529)}{
      \fill \p circle (1.35pt);
    }
    \node[phase] at (0,-1.82) {IIIC\\$120^\circ$ coplanar};
  \end{scope}
\end{tikzpicture}
\caption{\label{fig:geometry} Triple-$Q$ geometry and representative ordered states, drawn in TikZ using the notation and color conventions of Ref.~\cite{Jin2025}. (a) Three physically labeled ordering wave vectors at the distinct, symmetry-related $M$ points of the hexagonal Brillouin zone. (b)--(f) Fourier-amplitude configurations for phases I, IIB, IIIA, IIIB, and IIIC. Green, red, and blue consistently label the amplitudes at $\bm M_1$, $\bm M_2$, and $\bm M_3$, respectively. The central arrows show relative spin-space orientations; the mutually orthogonal IIIB frame is shown schematically in planar projection.}
\end{figure*}

\section{Model, symmetry, and convention}
\label{sec:model}

\subsection{Triple-$Q$ theory and fixed-lattice convention}
\label{subsec:convention}

We describe the magnetic order by three real $N$-component Fourier amplitudes, one for each $M$ point, and assemble them as columns of the $N\times3$ matrix
\begin{equation}
 \Phi=(\QQ_1\ \QQ_2\ \QQ_3),\qquad \QQ_l\in\mathbb R^N.
 \label{eq:Phi}
\end{equation}
This minimal description retains one magnetic mode at each ordering wave vector $\MM_l$, while neglecting additional sublattice form factors and modes belonging to other irreducible representations.
The corresponding long-wavelength spin density is
\begin{equation}
 \bm S(\bm r)=\sum_{l=1}^{3}\QQ_l(\bm r)
 e^{i\MM_l\cdot\bm r}+\text{c.c.}.
 \label{eq:spin_density}
\end{equation}
Because $2\MM_l$ is a reciprocal-lattice vector, each $M$ point is equivalent to its negative, and the amplitudes $\QQ_l$ may be chosen real.

To quartic order, the uniform free-energy density is
\begin{align}
 f_0={}&\alpha\sum_l|\QQ_l|^2
 +\beta_1\sum_l|\QQ_l|^4
 +\beta_2\sum_{l<m}|\QQ_l|^2|\QQ_m|^2 \nonumber\\
 &+\beta_3\sum_{l<m}(\QQ_l\cdot\QQ_m)^2.
 \label{eq:quartic}
\end{align}
Here $\alpha$ and $\beta_i$ are phenomenological coefficients.  We adopt the phase labels of Ref.~\cite{Jin2025}.  Depending on the quartic coefficients, the ordered minima comprise three phases for $N=3$: the single-$Q$ phase I, collinear triple-$Q$ phase IIIA, and orthogonal triple-$Q$ phase IIIB.  For $N=2$, they comprise four phases: phases I and IIIA, the orthogonal double-$Q$ phase IIB, and the $120^\circ$ triple-$Q$ phase IIIC.

The amplitudes $\QQ_l$ are physically distinguished by their associated momenta $\MM_l$.  Translations can change their signs, while point-group transformations can permute their labels; these actions relate distinct ordered states rather than gauge-equivalent descriptions of the same state.  States related by these transformations can belong to distinct crystalline domains, represented mathematically by different connected components of the ordered-state space.

By \emph{fixed lattice} we mean that the atomic lattice is treated as a prescribed, nondynamical background, with lattice dislocations and disclinations excluded.  To classify a free magnetic defect, we trace the order parameter along a closed path within one crystalline domain.  The magnetic order may wind, but the endpoint must coincide with the initial physical state, with neither a translation-induced sign change nor a point-group permutation of the $M$-point amplitudes.  A loop that closes only after a nontrivial crystalline transformation requires a domain wall and therefore does not represent a free internal defect.  A loop requiring a translational or rotational mismatch of the atomic lattice lies outside the fixed-lattice classification.  Magnetic textures and crystalline domain walls are thus retained as distinct objects.

This is a restriction on the classification, not an assumption that real crystals lack lattice defects.  Defects combining magnetic textures with lattice dislocations or disclinations require an enlarged configuration space and are not considered here.  At the other extreme, treating permutations of the $M$-point labels as gauge identifications would erase distinct crystalline domains and generate spurious internal defect charges.

Figure~\ref{fig:geometry} shows representative Fourier-amplitude configurations, while Table~\ref{tab:classification} summarizes their defect content, derived in Sec.~\ref{sec:classification}.

\subsection{Crystalline symmetry and parent group}

We now make the crystalline action on the physically labeled $M$-point amplitudes explicit.

Let $\bm a_1,\bm a_2$ be primitive Bravais-lattice vectors and $\bm b_1,\bm b_2$ the reciprocal basis, with $\bm b_i\cdot\bm a_j=2\pi\delta_{ij}$.  We choose the three $M$ points as
\[
 \MM_1=\bm b_1/2,\qquad
 \MM_2=\bm b_2/2,\qquad
 \MM_3=(\bm b_1+\bm b_2)/2,
\]
modulo reciprocal-lattice vectors.  Because $2\MM_l$ is reciprocal, translating by any Bravais vector $\bm a$ multiplies the corresponding amplitude by a sign,
\[
 \QQ_l\longmapsto\sigma_l(\bm a)\QQ_l,\qquad
 \sigma_l(\bm a)=e^{i\MM_l\cdot\bm a}=\pm1.
\]
Collecting these signs into $\bm\sigma(\bm a)=(\sigma_1,\sigma_2,\sigma_3)$, the translation acts on $\Phi$ as
\[
 T_{\bm a}:\Phi\longmapsto
 \Phi\,\operatorname{diag}[\bm\sigma(\bm a)].
\]
We call $\bm\sigma(\bm a)$ the translation sign pattern.

For the above choice of $M$ points, the primitive translations act as
\begin{align}
 T_{\bm a_1}:\Phi&\longmapsto
 \Phi\,\mathrm{diag}(-,+,-),\nonumber\\
 T_{\bm a_2}:\Phi&\longmapsto
 \Phi\,\mathrm{diag}(+,-,-).
 \label{eq:translations}
\end{align}
These transformations commute and each squares to the identity.  Translations by $2\bm a_1$ and $2\bm a_2$ act trivially.  The distinct translation actions therefore form the quotient by the common kernel generated by these two translations:
\begin{equation}
 \begin{split}
 \Vfour={}&\{(+,+,+),(-,+,-),\\
 &\hspace{2.6em}(+,-,-),(-,-,+)\}
 \simeq\ZZ_2\times\ZZ_2.
 \end{split}
 \label{eq:V4}
\end{equation}
Every allowed sign pattern satisfies $\sigma_1\sigma_2\sigma_3=+1$: the identity leaves all three amplitudes unchanged, whereas each nonidentity element reverses exactly two.  When the corresponding translations are broken, these operations relate distinct antiphase magnetic domains rather than gauge-equivalent descriptions.  The four domains and the primitive translations connecting them are shown in Fig.~\ref{fig:defect_atlas}(b).

The point-group action on the three $M$-point labels is $\Dthree$.  It permutes the labels and, correspondingly, the three nonidentity elements of $\Vfour$.  The effective finite crystalline group acting on the order parameters is therefore
\begin{equation}
 S=\Vfour\rtimes\Dthree.
 \label{eq:S}
\end{equation}
Let $\rho(s)$ be the $3\times3$ signed-permutation matrix representing $s\in S$ in label space.  The full internal symmetry is $\OO(N)$: proper rotations form its identity component, while improper transformations reverse internal orientation and may connect distinct chirality or frame-orientation sectors.  For the spin-isotropic theory, the parent group and its action are
\begin{align}
 G_N&=S\times\OO(N),&
 (s,R):\Phi&\longmapsto R\Phi\rho(s)^{-1}.
 \label{eq:group_action}
\end{align}
Because $S$ is discrete and the identity component of $\OO(N)$ is $\SO(N)$, the identity component of $G_N$ is
\begin{equation}
 G_0=\{e\}\times\SO(N).
 \label{eq:G0}
\end{equation}

A group element $g=(s,R)$ \emph{stabilizes} a state $\Phi_0$ if it leaves that point unchanged under the group action.  The crystalline and spin transformations may compensate each other,
\[
 R\Phi_0\rho(s)^{-1}=\Phi_0,
\]
even when neither transformation fixes $\Phi_0$ separately.  Here ``stabilizes'' means fixing a point in order-parameter space, not making the state energetically stable.

\section{Homotopy classification with a disconnected parent group}
\label{sec:homotopy}

\begin{figure*}[tb]
\centering
\begin{tikzpicture}[x=1cm,y=1cm,font=\small,
  note/.append style={font=\small}]
  \begin{scope}[shift={(0,0)}]
    \node[font=\bfseries] at (-0.40,2.30) {(a)};
    \node[font=\bfseries,text=qred] at (3.5,2.30)
      {projection used as stabilizer (incorrect)};
    \node[groupbox,text width=1.75cm] (hL) at (0.85,1.05)
      {combined pair\\$(s,R_s)$ fixes $\Phi_0$};
    \node[groupbox,text width=1.75cm,fill=qred!6] (pL) at (3.55,1.05)
      {projection\\forgets $s\neq e$};
    \node[groupbox,text width=2.15cm] (rL) at (6.45,1.05)
      {$R_s$ falsely appears\\to fix $\Phi_0$ alone};
    \draw[flow,qgray] (hL) -- (pL);
    \draw[flow,qred] (pL) -- (rL);

    \coordinate (iL) at (1.15,-0.65);
    \coordinate (rsL) at (5.80,-0.65);
    \fill (iL) circle (1.7pt);
    \fill[qred] (rsL) circle (1.7pt);
    \node[left=5pt] at (iL) {$I$};
    \node[right=5pt,text=qred] at (rsL) {$R_s$};
    \draw[flow,qblue] (iL) .. controls (2.55,0.10) and (4.35,0.10) .. (rsL)
      node[midway,above=4pt] {internal path};
    \draw[flow,qred,dashed] (rsL) .. controls (4.35,-1.55) and (2.45,-1.55) .. (iL);
    \node[text=qred] at (3.48,-1.62) {$s\neq e$: different crystalline state};
    \node[text=qred,font=\Large] at (6.75,-0.65) {$\times$};
  \end{scope}

  \begin{scope}[shift={(8.85,0)}]
    \node[font=\bfseries] at (0.0,2.30) {(b)};
    \node[font=\bfseries,text=qgreen] at (3.45,2.30)
      {connected stabilizer by intersection};
    \node[groupbox,text width=1.75cm] (hR) at (0.75,1.05) {$H\subset G_N$};
    \node[groupbox,text width=1.75cm,fill=qgreen!7] (pR) at (3.45,1.05)
      {retain only\\$s=e$};
    \node[groupbox,text width=1.85cm] (rR) at (6.15,1.05)
      {$K=H\cap G_0$};
    \draw[flow,qgray] (hR) -- (pR);
    \draw[flow,qgreen] (pR) -- (rR);

    \coordinate (cR) at (3.45,-0.65);
    \draw[line width=0.9pt,qgreen,flow]
      ($(cR)+(0:0.80)$) arc[start angle=0,end angle=335,radius=0.80];
    \fill ($(cR)+(0:0.80)$) circle (1.6pt);
    \node at (cR) {$I\to I$};
    \node[note,text=qgreen] at (3.45,-1.80)
      {closed within one fixed-lattice component};
  \end{scope}
\end{tikzpicture}
\caption{\label{fig:construction} Stabilizer construction, where $I$ denotes the identity spin transformation.  A pair $(s,R_s)$ may fix $\Phi_0$ although neither part does separately.  (a) Projection onto spin space is mathematically well defined, but using the projected group as the stabilizer drops $s\ne e$ and falsely identifies $R_s$ as an internal closure even though the endpoint differs by a nontrivial crystalline transformation.  (b) The connected stabilizer is instead obtained by intersection, which retains only trivial crystalline parts and ensures that loops close within one fixed-lattice component.}
\end{figure*}

The classification proceeds in three steps.  First, we construct the full symmetry orbit of a reference minimum $\Phi_0$.  Second, we isolate the connected component containing $\Phi_0$ that is accessible through continuous transformations on the fixed lattice.  Third, we compute the homotopy groups of that component.  The stabilizer, or isotropy subgroup, of $\Phi_0$ and its full ordered-state space are
\begin{align}
 H&=\{g\in G_N\mid g\Phi_0=\Phi_0\},&
 \mathcal M&=G_N/H.
 \label{eq:H_M}
\end{align}
Two elements of $G_N$ represent the same ordered state when they differ by right multiplication by an element of $H$, which yields the quotient $G_N/H$.

A continuous path in $\mathcal M$ beginning at $\Phi_0$ can be lifted to a path $g:[0,1]\to G_N$, $t\mapsto g(t)$, with $g(0)=e$.  Because the lift is continuous and begins at the identity, it remains in $G_0$.  Such a path is a loop based at $\Phi_0$ only if $g(1)\in H$; hence its endpoint must lie in
\begin{equation}
 K=H\cap G_0.
 \label{eq:K}
\end{equation}
The connected component containing $\Phi_0$ is therefore
\begin{equation}
 \mathcal M_0\simeq G_0/K.
 \label{eq:M0}
\end{equation}
The other components are labeled by the coset set
\begin{equation}
 \pi_0(\mathcal M)\simeq G_N/(G_0H).
 \label{eq:pi0}
\end{equation}
Because $G_0H$ need not be a normal subgroup of $G_N$, this quotient is generally only a set of domain sectors, not a group with a universal wall-composition law.

\paragraph{Why intersection, not projection.}
Under the natural identification $G_0\simeq\SO(N)$, the correct internal stabilizer and the subgroup obtained by projection are
\begin{align}
 K_{\mathrm{spin}}
 &=\{R\in\SO(N)\mid(e,R)\in H\},\nonumber\\
 K_{\mathrm{proj}}
 &=\mathrm{pr}_{\OO(N)}(H)\cap\SO(N).
 \label{eq:projection_error}
\end{align}
We always have $K_{\mathrm{spin}}\subseteq K_{\mathrm{proj}}$, but the inclusion is strict whenever $H$ contains a combined stabilizer $(s,R_s)$ with $s\ne e$ and $(e,R_s)\notin H$.  Quotienting by $K_{\mathrm{proj}}$ would then falsely close an internal path at
\[
 R_s\Phi_0=s^{-1}\Phi_0,
\]
even though this endpoint is a physically distinct crystalline-related state.  The correct connected manifold remains $G_0/(H\cap G_0)$, as illustrated in Fig.~\ref{fig:construction}.

Having isolated $\mathcal M_0=G_0/K$, we compute its homotopy groups using the universal cover $p:\widetilde G_0\to G_0$.  Let $\widetilde K=p^{-1}(K)$ and let $\widetilde K_0$ denote its identity component.  The long exact homotopy sequence for the fibration
\[
 \widetilde K\longrightarrow\widetilde G_0
 \longrightarrow\mathcal M_0
\]
and the facts that $\widetilde G_0$ is simply connected and $\pi_2(\widetilde G_0)=0$ give
\begin{align}
 \pi_1(\mathcal M_0)&\simeq\pi_0(\widetilde K),&
 \pi_2(\mathcal M_0)&\simeq\pi_1(\widetilde K_0).
 \label{eq:cover}
\end{align}
These formulas yield the three cases repeatedly used below.  For $G_0=\SO(3)$ and $K=\{e\}$, $\widetilde K=\{\pm I\}$, the kernel of $\SU(2)\to\SO(3)$, so $\pi_1(\mathcal M_0)=\ZZ_2$.  For $G_0=\SO(2)$ and $K=\{e\}$, the preimage of the identity under $\mathbb R\to\SO(2)$ is $2\pi\ZZ$, giving $\pi_1(\mathcal M_0)=\ZZ$.  Finally, for $G_0=\SO(3)$ and $K=\SO(2)$, $\widetilde K_0\simeq U(1)$, giving
\[
 \pi_2[\SO(3)/\SO(2)]=\pi_2(S^2)=\ZZ.
\]

In two spatial dimensions, $\pi_0$, $\pi_1$, and $\pi_2$ describe domain walls, point vortices, and nonsingular skyrmion textures, respectively.  In three dimensions, they describe surface domain walls, line vortices, and point defects such as hedgehogs.  The elastic theory in Sec.~\ref{sec:elasticity} is formulated for a single two-dimensional layer; the three-dimensional statements refer to a stacked extension with the same local ordered-state manifold and additional gradients along the stacking direction.  The $\pi_0$, $\pi_1$, and $\pi_2$ classification here does not exhaust the nonsingular textures possible in three dimensions, which may additionally be governed by higher homotopy groups such as $\pi_3(\mathcal M_0)$.

Vortex charges compose by concatenating loops in $\pi_1(\mathcal M_0)$.  They are Abelian when this group is commutative, so the combined charge is independent of order, and non-Abelian when the order of composition matters.  Thus the $\ZZ$ and $\ZZ_2$ vortices found here are Abelian, whereas a binary-polyhedral fundamental group such as $2T$ would produce non-Abelian vortex charges.

\paragraph{Graph-stabilizer lemma.}
A useful special case occurs when every element of a crystalline subgroup $S'\subset S$ is compensated by an internal transformation specified by a homomorphism $f:S'\to\OO(N)$.  The corresponding stabilizer is the graph
\begin{equation}
 H_f=\{(s,f(s))\mid s\in S'\}.
 \label{eq:graph_stabilizer}
\end{equation}
Because $f(e)=I$, the only graph element in $G_0$ is $(e,I)$, and hence
\[
 H_f\cap G_0=\{(e,I)\}.
\]
Choosing one representative of each coset in $S/S'$, every coset of $H_f$ has a unique representative with that crystalline part and an $\OO(N)$ coordinate.  Therefore
\begin{equation}
 (S\times\OO(N))/H_f
 \simeq\bigsqcup_{[s]\in S/S'}\OO(N).
 \label{eq:graph_quotient}
\end{equation}
Here $S/S'$ is a coset set; $S'$ need not be normal.  Phases IIIB, IIIC, and IIB are direct applications of this lemma.

The crucial point is that every nontrivial compensating graph element has a crystalline part $s\ne e$ and therefore lies outside $G_0$.  In the cyclic phase of a spin-2 Bose--Einstein condensate, by contrast, the parent gauge--spin group is connected.  Combined gauge--spin stabilizers then remain inside its identity component and can legitimately generate non-Abelian vortices~\cite{Kobayashi2009,KawaguchiUeda2012}.

Accordingly, each phase calculation consists of determining $H$, forming $K=H\cap G_0$, identifying the connected components, and applying Eq.~\eqref{eq:cover} to compute $\pi_1$ and $\pi_2$.

\section{Classification of the seven stable phases}
\label{sec:classification}

\begin{figure*}[tb]
\centering
\begin{tikzpicture}[x=1cm,y=1cm,font=\small,
  note/.append style={font=\small}]
  \begin{scope}[shift={(0,0)}]
    \node[font=\bfseries] at (-0.15,1.55) {(a)};
    \node[font=\bfseries] at (2.05,1.55) {$Q$-selection sectors};
    \foreach \x/\lab/\col in {0.65/1/qgreen,2.05/2/qred,3.45/3/qblue}{
      \node[draw=\col,rounded corners=2pt,fill=\col!7,
        minimum width=1.05cm,minimum height=0.72cm]
        at (\x,0.40) {$\alpha=\lab$};
    }
    \node[note] at (2.05,-0.38)
      {active $Q_\alpha$ (I); missing $Q_\alpha$ (IIB)};
  \end{scope}

  \begin{scope}[shift={(5.35,0)}]
    \node[font=\bfseries] at (-0.15,1.55) {(b)};
    \node[font=\bfseries] at (2.05,1.55) {$\Vfour$ antiphase sectors};
    \node[draw=qgray,rounded corners=2pt,fill=qgray!5,align=center,
      minimum width=1.75cm,minimum height=0.72cm] (v0)
      at (0.85,0.70) {$e$\\$(+,+,+)$};
    \node[draw=qgray,rounded corners=2pt,fill=qgray!5,align=center,
      minimum width=1.75cm,minimum height=0.72cm] (v1)
      at (3.25,0.70) {$T_{\bm a_1}$\\$(-,+,-)$};
    \node[draw=qgray,rounded corners=2pt,fill=qgray!5,align=center,
      minimum width=1.75cm,minimum height=0.72cm] (v2)
      at (0.85,-0.38) {$T_{\bm a_2}$\\$(+,-,-)$};
    \node[draw=qgray,rounded corners=2pt,fill=qgray!5,align=center,
      minimum width=1.75cm,minimum height=0.72cm] (v3)
      at (3.25,-0.38) {$T_{\bm a_1+\bm a_2}$\\$(-,-,+)$};
    \draw[flow,qred] (v0) -- (v1);
    \draw[flow,qblue] (v0) -- (v2);
    \draw[flow,qblue] (v1) -- (v3);
    \draw[flow,qred] (v2) -- (v3);
    \node[note] at (2.05,-1.05)
      {\textcolor{qred}{$T_{\bm a_1}$}: horizontal;\quad
       \textcolor{qblue}{$T_{\bm a_2}$}: vertical};
  \end{scope}

  \begin{scope}[shift={(10.75,0)}]
    \node[font=\bfseries] at (-0.15,1.55) {(c)};
    \node[font=\bfseries] at (2.05,1.55)
      {frame/chirality sectors};
    \filldraw[fill=qgreen!8,draw=qgreen,line width=0.8pt]
      (0.95,0.35) circle (0.68);
    \filldraw[fill=qorange!9,draw=qorange,line width=0.8pt]
      (3.15,0.35) circle (0.68);
    \node at (0.95,0.35) {$\mathcal M_{+}$};
    \node at (3.15,0.35) {$\mathcal M_{-}$};
    \draw[qred,line width=0.95pt,decorate,
      decoration={snake,amplitude=0.65mm,segment length=2.2mm}]
      (2.05,-0.38) -- (2.05,1.08);
    \node[note,text=qred] at (2.05,-0.72) {discrete-domain wall};
  \end{scope}

  \begin{scope}[shift={(0,-3.85)}]
    \node[font=\bfseries] at (-0.15,1.55) {(d)};
    \node[font=\bfseries] at (2.05,1.55)
      {$S^1$ integer vortex};
    \coordinate (c) at (2.05,0.25);
    \draw[qgray!60,dashed] (c) circle (0.92);
    \filldraw[fill=white,draw=qgray,line width=0.8pt] (c) circle (0.11);
    \foreach \a in {0,45,...,315}{
      \draw[-{Latex[length=1.45mm,width=0.95mm]},qblue,line width=0.75pt]
        ($(c)+(\a:0.60)$) -- ++(\a:0.30);
    }
    \draw[flow,qblue]
      ($(c)+(8:1.13)$) arc[start angle=8,end angle=338,radius=1.13];
    \node[note,text=qblue] at (2.05,-1.04)
      {$\theta:0\rightarrow2\pi$};
  \end{scope}

  \begin{scope}[shift={(5.35,-3.85)}]
    \node[font=\bfseries] at (-0.15,1.55) {(e)};
    \node[font=\bfseries] at (2.05,1.55)
      {$\SO(3)$ $\ZZ_2$ frame vortex};
    \coordinate (c) at (2.05,0.25);
    \draw[qgray!60,dashed] (c) circle (0.92);
    \filldraw[fill=white,draw=qgray,line width=0.8pt] (c) circle (0.11);
    \draw[flow,qgray]
      ($(c)+(8:1.13)$) arc[start angle=8,end angle=338,radius=1.13];
    \foreach \a in {0,60,...,300}{
      \coordinate (p) at ($(c)+(\a:0.70)$);
      \begin{scope}[shift={(p)},rotate=\a]
        \draw[-{Latex[length=1.15mm,width=0.8mm]},qred,line width=0.65pt]
          (0,0) -- (0.25,0);
        \draw[-{Latex[length=1.15mm,width=0.8mm]},qblue,line width=0.65pt]
          (0,0) -- (0,0.25);
        \filldraw[fill=white,draw=qgreen,line width=0.6pt]
          (0,0) circle (0.058);
        \fill[qgreen] (0,0) circle (0.018);
      \end{scope}
    }
    \node[note,text=qgray] at (2.05,-1.04)
      {$\widetilde R:I\rightarrow-I$ in $\SU(2)$};
  \end{scope}

  \begin{scope}[shift={(10.75,-3.85)}]
    \node[font=\bfseries] at (-0.15,1.55) {(f)};
    \node[font=\bfseries] at (2.05,1.55)
      {$S^2$ skyrmion/hedgehog};
    \coordinate (c) at (2.05,0.25);
    \filldraw[fill=qgreen!4,draw=qgray,line width=0.8pt]
      (c) circle (0.92);
    \foreach \a in {0,45,...,315}{
      \coordinate (po) at ($(c)+(\a:0.72)$);
      \filldraw[fill=white,draw=qgreen,line width=0.6pt]
        (po) circle (0.058);
      \fill[qgreen] (po) circle (0.018);
      \coordinate (pm) at ($(c)+(\a:0.39)$);
      \draw[-{Latex[length=1.25mm,width=0.85mm]},qblue,line width=0.7pt]
        (pm) -- ++(\a:0.25);
    }
    \filldraw[fill=white,draw=qorange,line width=0.75pt]
      (c) circle (0.09);
    \draw[qorange,line width=0.75pt]
      ($(c)+(-0.045,-0.045)$) -- ($(c)+(0.045,0.045)$)
      ($(c)+(-0.045,0.045)$) -- ($(c)+(0.045,-0.045)$);
    \node[note,text=qgray] at (2.05,-1.04)
      {$Q_{\rm sk}=\pm1$ in 2D; elementary hedgehog in 3D};
  \end{scope}
\end{tikzpicture}
\caption{\label{fig:defect_atlas} Defect atlas keyed to the final column of Table~\ref{tab:classification}.  (a) Three choices of active $Q$ in phase I or missing $Q$ in phase IIB.  (b) The four translation-related $\Vfour$ antiphase sectors; red and blue arrows apply $T_{\bm a_1}$ and $T_{\bm a_2}$.  (c) Two components distinguished by frame orientation in phase IIB, scalar chirality in phase IIIB, or vector chirality in phase IIIC; $\mathcal M_+$ and $\mathcal M_-$ are each homeomorphic to the phase-specific connected manifold $\mathcal M_0$ listed in Table~\ref{tab:classification}.  (d) The integer $2\pi$ generator of an $S^1$ component.  (e) The nontrivial $\ZZ_2$ frame loop in $\SO(3)$, which lifts from $I$ to $-I$ in $\SU(2)$.  Red and blue arrows denote two frame axes and the green dot the third.  (f) An $S^2$ texture, representing a skyrmion in two dimensions or an elementary hedgehog in three dimensions.}
\end{figure*}
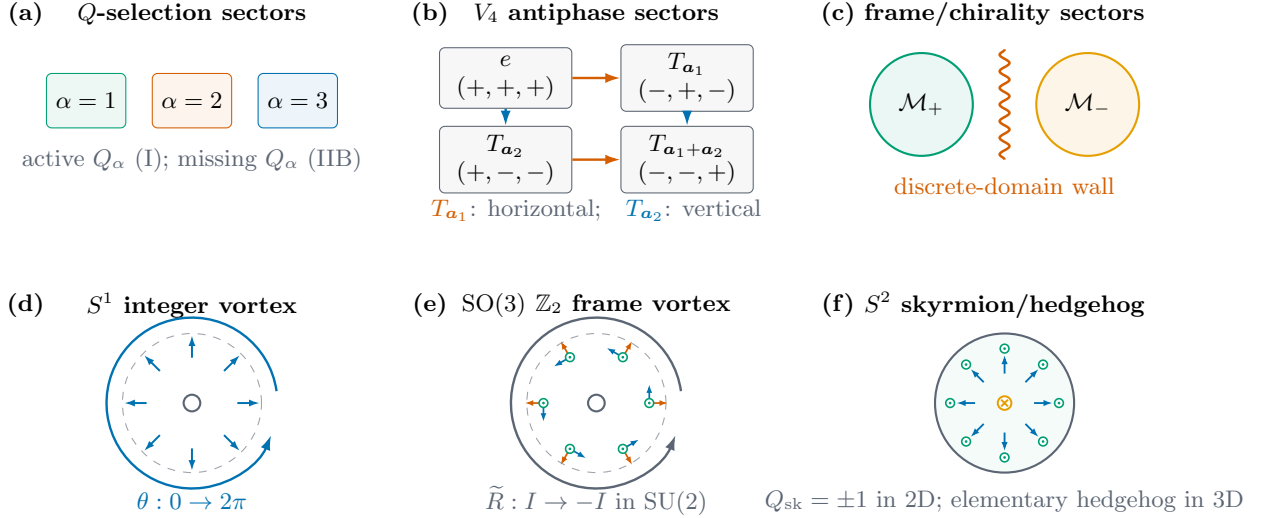

We now apply the construction of Sec.~\ref{sec:homotopy} to the seven stable phases.  Let $Q_0>0$ denote the equilibrium amplitude, let $\{\hat{\bm e}_i\}$ be an orthonormal internal basis, and, for $N=2$, define $\hat{\bm e}_\phi=(\cos\phi,\sin\phi)$.  Table~\ref{tab:classification} summarizes the resulting ordered-state manifolds and defect classifications.  The groups $\pi_1$ and $\pi_2$ are evaluated within a chosen connected component, while the disconnected domain sectors are displayed explicitly in the full manifold.  The final column identifies the corresponding building blocks in the defect atlas of Fig.~\ref{fig:defect_atlas}.

\begin{table*}[tb]
\caption{\label{tab:classification} Fixed-lattice classification of the seven stable phases.  The symbol $\bigsqcup_nY$ denotes $n$ copies of $Y$, while $\bigsqcup_{x\in X}Y$ denotes one copy for each element of the component set $X$.  The groups $\pi_1$ and $\pi_2$ are based at a point in one connected component; disconnected labels identify physical domain types but do not by themselves define a universal wall-composition law.  The final column refers to panels of Fig.~\ref{fig:defect_atlas}.}
\small
\renewcommand{\arraystretch}{1.25}
\setlength{\tabcolsep}{2.5pt}
\begin{tabular*}{\textwidth}{@{\extracolsep{\fill}}p{0.075\textwidth}p{0.18\textwidth}p{0.18\textwidth}
 p{0.042\textwidth}p{0.042\textwidth}p{0.29\textwidth}p{0.075\textwidth}@{}}
\toprule
Phase & Reference $\Phi_0$ in units of $Q_0$ & Full manifold $\mathcal M$ &
$\pi_1$ & $\pi_2$ & Defect content & Atlas panels \\
\midrule
I$_{\OO(3)}$ &
$(\hat{\bm e}_1\ \bm0\ \bm0)$ &
$\bigsqcup_{3}S^2$ &
$0$ & $\ZZ$ &
Three $Q$-selection sectors; skyrmions in 2D or hedgehogs in 3D &
a, f \\
IIIA$_{\OO(3)}$ &
$\hat{\bm e}_1(1,1,1)$ &
$\bigsqcup_{\Vfour}S^2$ &
$0$ & $\ZZ$ &
$\Vfour$ antiphase walls; skyrmions in 2D or hedgehogs in 3D &
b, f \\
IIIB$_{\OO(3)}$ &
$I_{3\times3}$ &
$\OO(3)=\bigsqcup_2\SO(3)$ &
$\ZZ_2$ & $0$ &
Scalar-chirality walls; Abelian $\ZZ_2$ frame vortices &
c, e \\
\midrule
I$_{\OO(2)}$ &
$(\hat{\bm e}_1\ \bm0\ \bm0)$ &
$\bigsqcup_{3}S^1$ &
$\ZZ$ & $0$ &
Three $Q$-selection sectors; integer $2\pi$ vortices &
a, d \\
IIB$_{\OO(2)}$ &
$(\hat{\bm e}_1\ \hat{\bm e}_2\ \bm0)$ &
$\bigsqcup_{3}\OO(2)=\bigsqcup_{6}S^1$ &
$\ZZ$ & $0$ &
Missing-$Q$ and frame-orientation walls; integer $2\pi$ frame vortices &
a, c, d \\
IIIA$_{\OO(2)}$ &
$\hat{\bm e}_1(1,1,1)$ &
$\bigsqcup_{\Vfour}S^1$ &
$\ZZ$ & $0$ &
$\Vfour$ antiphase walls; integer $2\pi$ vortices &
b, d \\
IIIC$_{\OO(2)}$ &
$(\hat{\bm e}_0\ \hat{\bm e}_{2\pi/3}\ \hat{\bm e}_{4\pi/3})$ &
$\Vfour\times\OO(2)=\bigsqcup_{8}S^1$ &
$\ZZ$ & $0$ &
Antiphase and vector-chirality walls; integer $2\pi$ vortices &
b, c, d \\
\bottomrule
\end{tabular*}
\end{table*}

We first treat phases I and IIIA, whose continuous stabilizers can be read off directly.  We then consider the graph-stabilized noncollinear phases IIIB, IIIC, and IIB.  Phase IIIB is presented first because it most directly exposes the distinction between intersection and projection.

\subsection{Single-$Q$ phase I}

Take as reference state
\begin{equation}
 \Phi_0=Q_0(\hat{\bm e}_1\ \bm0\ \bm0).
 \label{eq:I_reference}
\end{equation}
For $N=3$, proper rotations about $\hat{\bm e}_1$ form $K\simeq\SO(2)$, giving the connected manifold $\SO(3)/\SO(2)\simeq S^2$.  For $N=2$, only the identity proper rotation fixes $\hat{\bm e}_1$, so $K=\{(e,I)\}$ and the connected manifold is $\SO(2)\simeq S^1$.

The $\Dthree$ point-group orbit gives three choices of active $M$ point.  Translations introduce no additional antiphase sectors: a translation either leaves the nonzero Fourier amplitude unchanged or reverses its sign.  The two signs are connected by a proper $\pi$ spin rotation, while the translation action on the two vanishing amplitudes has no physical effect.  Therefore,
\begin{equation}
 \mathcal M_{\mathrm{I},\OO(3)}=\bigsqcup_3S^2,\qquad
 \mathcal M_{\mathrm{I},\OO(2)}=\bigsqcup_3S^1.
 \label{eq:I_manifolds}
\end{equation}
The three components represent $Q$-selection domains.  Each $\OO(3)$ component supports $S^2$ skyrmion or hedgehog textures but no free vortex, whereas each $\OO(2)$ component supports integer $2\pi$ vortices.

\subsection{Collinear triple-$Q$ phase IIIA}

Take $\Phi_0=Q_0\hat{\bm e}_1(1,1,1)$.  Every permutation in $\Dthree$ leaves this state unchanged and therefore belongs to $H$.  By contrast, each nonidentity element of $\Vfour$ changes the relative signs of the three Fourier amplitudes and generates a distinct antiphase sector.  The continuous stabilizer $K$ is $\SO(2)$ for $N=3$ and trivial for $N=2$.  Hence
\begin{equation}
 \mathcal M_{\mathrm{IIIA},\OO(3)}
 =\bigsqcup_{\Vfour}S^2,\qquad
 \mathcal M_{\mathrm{IIIA},\OO(2)}
 =\bigsqcup_{\Vfour}S^1.
 \label{eq:IIIA_manifolds}
\end{equation}
For the planar phase, write $\QQ_l=Q_0\tau_l\hat{\bm e}_\theta$, with $\tau_l=\pm1$.  The simultaneous replacement
\[
 (\theta,\bm\tau)\sim(\theta+\pi,-\bm\tau)
\]
leaves the physical Fourier amplitudes unchanged.  Choosing the representative satisfying $\tau_1\tau_2\tau_3=+1$ gives the four $\Vfour$ sign patterns.  Importantly, this parametrization redundancy does not identify $(\theta,\bm\tau)$ with $(\theta+\pi,\bm\tau)$.  A loop within a fixed antiphase sector therefore closes only after a $2\pi$ winding.

The four components are $\Vfour$ antiphase domains.  Their $\OO(3)$ versions support $S^2$ textures, whereas their $\OO(2)$ versions support integer $2\pi$ vortices.

\subsection{Orthogonal triple-$Q$ phase IIIB}
\label{subsec:IIIB}

Take the orthonormal reference frame
\begin{equation}
 \Phi_0=Q_0 I_{3\times3}.
 \label{eq:IIIB_reference}
\end{equation}
Because $\Phi_0$ has full rank, the stabilizer condition uniquely fixes the spin transformation accompanying each $s\in S$ to be $\rho(s)$.  Thus
\begin{equation}
 H=\{(s,\rho(s))\mid s\in S\}.
 \label{eq:IIIB_H}
\end{equation}
Its proper-spin projection is the tetrahedral rotation group $T$:
\begin{equation}
 \mathrm{pr}_{\OO(3)}(H)\cap\SO(3)\simeq T.
 \label{eq:tetrahedral_projection}
\end{equation}
These tetrahedral rotations are not pure internal stabilizers: every nonidentity rotation is tied to a nontrivial crystalline operation.  Hence
\begin{equation}
 K=H\cap G_0=\{(e,I)\},\qquad
 \mathcal M_0\simeq\SO(3),
 \label{eq:IIIB_K}
\end{equation}
rather than $\SO(3)/T$.

The full quotient is isomorphic to $\OO(3)$.  Explicitly, the map
\begin{equation}
 [(s,R)]\longmapsto R\rho(s)^{-1}
 \label{eq:IIIB_isomorphism}
\end{equation}
is constant on cosets of $H$ and is bijective.  Therefore
\begin{equation}
 \mathcal M\simeq\OO(3)
 =\SO(3)\sqcup\bigl[\OO(3)\setminus\SO(3)\bigr].
 \label{eq:IIIB_M}
\end{equation}
The two components are distinguished by the normalized scalar chirality
\begin{equation}
 \chi=\frac{\QQ_1\cdot(\QQ_2\times\QQ_3)}{Q_0^3}
 =\det(\Phi/Q_0)=\pm1.
 \label{eq:IIIB_chirality}
\end{equation}
Consequently,
\begin{align}
 \pi_0(\mathcal M)&=\{\chi=+1,\chi=-1\},\nonumber\\
 \pi_1(\mathcal M_0)&=\ZZ_2,\qquad
 \pi_2(\mathcal M_0)=0.
 \label{eq:IIIB_homotopy}
\end{align}
The nontrivial free defect is therefore an Abelian $\ZZ_2$ frame vortex: a point vortex in two dimensions or a line vortex in three.  It belongs to the same $\pi_1[\SO(3)]=\ZZ_2$ class as the classic vortex of the triangular Heisenberg antiferromagnet~\cite{KawamuraMiyashita1984,Kawamura2011}.  Incorrectly replacing $K$ by $T$ would instead yield $\SO(3)/T$ and the non-Abelian binary-tetrahedral fundamental group $2T$; that quotient does not describe the fixed-lattice manifold of physically labeled Fourier fields.

\subsection{$120^\circ$ triple-$Q$ phase IIIC}
\label{subsec:IIIC}

Take
\begin{equation}
 \Phi_0=Q_0(\hat{\bm e}_0\
 \hat{\bm e}_{2\pi/3}\ \hat{\bm e}_{4\pi/3}).
 \label{eq:IIIC_reference}
\end{equation}
Each permutation $d\in\Dthree$ has a unique compensating planar transformation $r_d$.  The stabilizer is therefore the graph
\begin{align}
 H&=\{(d,r_d)\mid d\in\Dthree\}\simeq\Dthree,\nonumber\\
 K&=H\cap G_0=\{(e,I)\}.
 \label{eq:IIIC_HK}
\end{align}
No nonidentity translation belongs to $H$.  The graph-stabilizer construction then gives four $\Vfour$ antiphase sectors, each containing the two components of $\OO(2)$:
\begin{equation}
 \mathcal M=\Vfour\times\OO(2)=\bigsqcup_8S^1.
 \label{eq:IIIC_M}
\end{equation}
The two orientation components are distinguished by the vector chirality
\begin{equation}
 \kappa=\operatorname{sgn}\!\left[
 \hat{\bm z}\cdot\sum_{l=1}^{3}
 \QQ_l\times\QQ_{l+1}\right],
 \qquad \QQ_4\equiv\QQ_1,
 \label{eq:IIIC_chirality}
\end{equation}
which is reversed by a pure internal or crystalline reflection, whereas their compensated combination belongs to $H$.

To display the antiphase and chirality labels simultaneously, let $\tau_l=(-1)^{n_l}$, with $n_l\in\ZZ_2$, and define
\begin{equation}
 \lambda=\operatorname{sgn}
 \bigl(\tau_1\tau_2+\tau_2\tau_3+\tau_3\tau_1\bigr).
 \label{eq:IIIC_lambda}
\end{equation}
Thus $\lambda=+1$ when all three signs agree and $\lambda=-1$ otherwise.  A convenient parametrization is
\begin{align}
 \phi_l&=\theta+\lambda\kappa\frac{2\pi(l-1)}{3}+\pi n_l,\nonumber\\
 \QQ_l&=Q_0\hat{\bm e}_{\phi_l},
 \qquad \kappa=\pm1,\quad n_l\in\ZZ_2,
 \label{eq:IIIC_param}
\end{align}
where the factor $\lambda$ ensures that $\kappa$ remains the physical vector-chirality label in every antiphase sector.

The simultaneous replacement
\[
 \theta\mapsto\theta+\pi,\qquad n_l\mapsto n_l+1
\]
leaves all three Fourier amplitudes unchanged, reducing the eight sign triples to four antiphase classes.  In contrast,
\begin{equation}
 (\theta,\kappa)\sim(\theta+\pi,-\kappa)
 \label{eq:false_identification}
\end{equation}
is not a stabilizer identification.  Every component therefore has $\pi_1=\ZZ$ with a $2\pi$ generator.  A $\pi$ winding can close only across a wall that changes the chirality or antiphase sector.

\subsection{Orthogonal double-$Q$ phase IIB}

Take
\begin{equation}
 \Phi_0=Q_0(\hat{\bm e}_1\ \hat{\bm e}_2\ \bm0).
 \label{eq:IIB_reference}
\end{equation}
The two occupied columns form an orthonormal planar frame.  Their eight signed permutations constitute a subgroup $\mathcal P_2\subset\OO(2)$, and each is uniquely compensated by a crystalline operation that preserves the missing $M$-point label.  The stabilizer is therefore an eight-element graph, given explicitly in Appendix~\ref{app:IIB}.  Only its identity element has a trivial crystalline part, so
\[
 K=H\cap G_0=\{(e,I)\}.
\]

There are three choices of missing $M$ point.  For each choice, the frame spans $\OO(2)$, whose two connected components are distinguished by its orientation.  Hence
\begin{equation}
 \mathcal M_{\mathrm{IIB},\OO(2)}
 =\bigsqcup_3\OO(2)=\bigsqcup_6S^1.
 \label{eq:IIB_manifold}
\end{equation}
The phase therefore supports missing-$Q$ walls, frame-orientation walls, and integer $2\pi$ frame vortices.  This completes the seven entries in Table~\ref{tab:classification}; in particular, every connected component of every planar phase has a $2\pi$ vortex generator.

Across all seven phases, discrete crystalline labels generate domain sectors but do not shorten loops within a fixed component.  The connected components are $S^2$, $S^1$, or $\SO(3)$, supporting respectively $S^2$ textures, integer $2\pi$ vortices, or Abelian $\ZZ_2$ frame vortices.  Half windings require a domain wall, and no non-Abelian free vortex occurs in the fixed-lattice theory.

\section{Symmetry-constrained elastic sector}
\label{sec:elasticity}

Homotopy determines which defects are allowed, but their profiles and energies are governed by the elastic energy of the slowly varying Fourier amplitudes $\QQ_l(\bm r)$ within the Ginzburg--Landau framework of Ref.~\cite{Jin2025}.  We first impose translation symmetry and then the point group.  The most general quadratic gradient term that is scalar under $\OO(N)$ is
\begin{equation}
 f_{\nabla}^{(2)}
 =\sum_{l,m=1}^{3}\sum_{i,j=x,y}
 K_{lm}^{ij}(\partial_i\QQ_l)\cdot(\partial_j\QQ_m).
 \label{eq:general_gradient}
\end{equation}
Under a lattice translation, $\QQ_l$ acquires the phase associated with $\MM_l$.  A bilinear with labels $l$ and $m$ is therefore invariant only if $\MM_l+\MM_m$ is a reciprocal-lattice vector.  This condition holds for $l=m$, because $2\MM_l$ is reciprocal, but fails for $l\ne m$, because $\MM_l+\MM_m\equiv\MM_n$ with nonzero $\MM_n$.  Equivalently, every cross-label bilinear changes sign under at least one element of $\Vfour$.  Hence
\begin{equation}
 K_{lm}^{ij}=0\qquad(l\ne m).
 \label{eq:no_cross_gradient}
\end{equation}
Translation symmetry thus reduces the problem to the same-label sector; the point group will further reduce that sector to two independent stiffnesses.

To impose the point-group constraints, define $t_{l,ij}=(\partial_i\QQ_l)\cdot(\partial_j\QQ_l)$ and order the nine same-label bilinears as
\begin{equation}
 \bm t=(t_{1,xx},t_{2,xx},t_{3,xx},
 t_{1,xy},t_{2,xy},t_{3,xy},
 t_{1,yy},t_{2,yy},t_{3,yy}).
 \label{eq:gradient_basis}
\end{equation}
Writing the same-label quadratic form as $f_{\nabla}^{(2)}=\bm c\cdot\bm t$, choose a mirror $A$ and a $120^\circ$ rotation $D$ as generators of $\Dthree$.  Their actions on labels and spatial derivatives are
\begin{align}
 A:\quad
 (\QQ_1,\QQ_2,\QQ_3)&\mapsto(\QQ_1,\QQ_3,\QQ_2),
 \nonumber\\
 \partial_x&\mapsto-\partial_x,\qquad
 \partial_y\mapsto\partial_y,
 \label{eq:gradient_A}\\
 D:\quad
 (\QQ_1,\QQ_2,\QQ_3)&\mapsto(\QQ_2,\QQ_3,\QQ_1),
 \nonumber\\
 \partial_x&\mapsto
 \frac12\partial_x+\frac{\sqrt3}{2}\partial_y,
 \nonumber\\
 \partial_y&\mapsto
 -\frac{\sqrt3}{2}\partial_x+\frac12\partial_y.
 \label{eq:gradient_D}
\end{align}
Let $\mathsf A$ and $\mathsf D$ denote the corresponding induced linear maps on the coefficient vector $\bm c$.
Invariance requires
\begin{equation}
 (\mathsf A-I)\bm c=(\mathsf D-I)\bm c=0.
 \label{eq:gradient_nullspace}
\end{equation}
Appendix~\ref{app:gradient} constructs these maps and shows that their common kernel is two dimensional.  To express its invariant basis, let $\psi_l$ be the polar angle of $\MM_l$ in the convention of Fig.~\ref{fig:geometry}(a):
\begin{equation}
 (\psi_1,\psi_2,\psi_3)
 =\left(\frac{\pi}{2},\frac{7\pi}{6},\frac{11\pi}{6}\right).
 \label{eq:psi}
\end{equation}
A convenient basis is
\begin{align}
 \mathcal{I}_1={}&\sum_{l=1}^{3}|\nabla\QQ_l|^2,
 \label{eq:I1}\\
 \mathcal{I}_2={}&\sum_{l=1}^{3}\bigl[
 -\cos(2\psi_l)(|\partial_x\QQ_l|^2-|\partial_y\QQ_l|^2)
 \nonumber\\
 &\hspace{3.7em}
 +2\sin(2\psi_l)\,
 \partial_x\QQ_l\cdot\partial_y\QQ_l\bigr],
 \label{eq:I2}
\end{align}
where $\mathcal I_1$ is the isotropic invariant and $\mathcal I_2$ is the label-locked invariant that ties the spatial stiffness tensor of each Fourier field to its $M_l$ direction.
Thus
\begin{equation}
 f_{\nabla}^{(2)}=\frac{\rho}{2}\mathcal{I}_1+\frac{\eta}{2}\mathcal{I}_2.
 \label{eq:two_stiffness}
\end{equation}
This is the complete quadratic two-derivative elastic energy allowed by the ideal symmetry.
The label-locked coefficient $\eta$ describes spatial anisotropy of an otherwise spin-rotation-invariant stiffness and should not be confused with spin--orbit-induced anisotropic exchange.  Spin--orbit coupling generates spin-space tensors of two distinct types.  Symmetric anisotropy, including single-ion and symmetric exchange anisotropy, can modify both the uniform free energy $f_0$ and the two-derivative elastic coefficients.  Antisymmetric Dzyaloshinskii--Moriya exchange instead generates symmetry-allowed Lifshitz invariants linear in gradients~\cite{Dzyaloshinsky1958,Moriya1960}.  Both effects reduce the independent $\OO(N)$ symmetry and require the ordered-state manifolds to be reconsidered.
The anisotropic invariant $\mathcal I_2$ is mirror even and therefore does not select either scalar or vector chirality.  It changes defect energetics but not the ordered-state manifold or its homotopy groups.  Writing
$f_{\nabla}^{(2)}=\frac12\sum_l(\partial_i\QQ_l)\cdot(K_l)_{ij}(\partial_j\QQ_l)$, the stiffness matrix for label $l$ is
\begin{equation}
 K_l=
 \begin{pmatrix}
 \rho-\eta\cos2\psi_l & \eta\sin2\psi_l\\
 \eta\sin2\psi_l & \rho+\eta\cos2\psi_l
 \end{pmatrix},
 \label{eq:Kl}
\end{equation}
with principal stiffnesses $\rho\pm\eta$.  Positive definiteness therefore requires $\rho>|\eta|$.

\begin{figure*}[tb]
\centering
\begin{tikzpicture}[x=1cm,y=1cm,font=\small,
  note/.append style={font=\footnotesize}]
  \begin{scope}[shift={(0.10,0)}]
    \node[font=\bfseries] at (-0.20,1.55) {(a)};
    \node[font=\bfseries] at (2.05,1.55)
      {label-locked stiffness};
    \foreach \x/\ang/\col/\lab/\psilab in {
      0.55/0/qgreen/1/{\pi/2},
      2.05/60/qred/2/{7\pi/6},
      3.55/120/qblue/3/{11\pi/6}}{
      \begin{scope}[shift={(\x,0.20)},rotate=\ang]
        \filldraw[fill=\col!8,draw=\col,line width=0.8pt]
          (0,0) ellipse[x radius=0.42,y radius=0.72];
        \draw[\col!55,dashed,line width=0.55pt]
          (-0.58,0) -- (0.58,0);
        \draw[\col!55,dashed,line width=0.55pt]
          (0,-0.84) -- (0,0.84);
      \end{scope}
      \node[text=\col] at (\x,0.20) {$K_{\lab}$};
      \node[note,text=\col] at (\x,-0.60)
        {$\psilab$};
    }
  \end{scope}

  \begin{scope}[shift={(5.25,0)}]
    \node[font=\bfseries] at (-0.20,1.55) {(b)};
    \node[font=\bfseries] at (2.10,1.55)
      {equal contributions};
    \foreach \x/\ang/\col in {
      0.30/0/qgreen,
      1.40/60/qred,
      2.50/120/qblue}{
      \begin{scope}[shift={(\x,0.20)},rotate=\ang]
        \filldraw[fill=\col!8,draw=\col,line width=0.75pt]
          (0,0) ellipse[x radius=0.27,y radius=0.47];
      \end{scope}
    }
    \node at (0.85,0.20) {$+$};
    \node at (1.95,0.20) {$+$};
    \draw[flow,qgray] (2.95,0.20) -- (3.55,0.20);
    \filldraw[fill=qgray!5,draw=qgray,line width=0.9pt]
      (4.15,0.20) circle (0.55);
    \node at (4.15,0.20) {$3\rho\mathbf 1_2$};
    \node[note] at (2.15,-0.62)
      {$K_1+K_2+K_3=3\rho\mathbf 1_2$};
  \end{scope}

  \begin{scope}[shift={(10.90,0)}]
    \node[font=\bfseries] at (-0.20,1.55) {(c)};
    \node[font=\bfseries] at (2.10,1.55)
      {unequal contributions};
    \begin{scope}[shift={(0.35,0.20)},rotate=0]
      \filldraw[fill=qgreen!12,draw=qgreen,line width=1.0pt]
        (0,0) ellipse[x radius=0.32,y radius=0.56];
    \end{scope}
    \begin{scope}[shift={(1.35,0.20)},rotate=60]
      \filldraw[fill=qred!6,draw=qred!65,line width=0.65pt]
        (0,0) ellipse[x radius=0.22,y radius=0.38];
    \end{scope}
    \begin{scope}[shift={(2.25,0.20)},rotate=120]
      \filldraw[fill=qblue!6,draw=qblue!65,line width=0.65pt]
        (0,0) ellipse[x radius=0.20,y radius=0.35];
    \end{scope}
    \draw[flow,qgray] (2.72,0.20) -- (3.25,0.20);
    \begin{scope}[shift={(3.90,0.20)},rotate=8]
      \filldraw[fill=qorange!8,draw=qorange,line width=0.9pt]
        (0,0) ellipse[x radius=0.43,y radius=0.68];
    \end{scope}
    \node[note] at (2.10,-0.62)
      {$\sum_l w_lK_l\not\propto\mathbf 1_2$};
  \end{scope}
\end{tikzpicture}
\caption{\label{fig:elasticity} Stiffness geometry. (a) Label-locked constant-energy ellipses for $\eta>0$; their short and long axes correspond to $\rho+\eta$ and $\rho-\eta$ and interchange for $\eta<0$. (b) Equal derivative contributions from the three labels give $3\rho\mathbf 1_2$. (c) Unequal derivative contributions retain anisotropy in single- and double-$Q$ phases, generic phase-IIIB frame textures, and near cores or walls.}
\end{figure*}
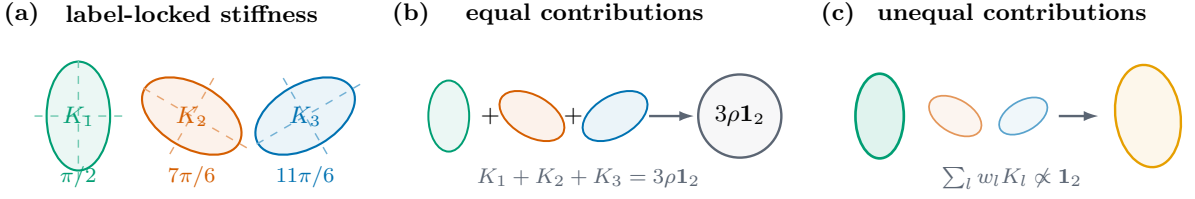

Although each $K_l$ is anisotropic, the three tensors satisfy
\begin{equation}
 \sum_{l=1}^{3}K_l=3\rho\,\mathbf{1}_2,
 \label{eq:anisotropy_cancellation}
\end{equation}
so the $\eta$ contribution cancels whenever the three labels have identical derivative tensors.  This occurs for the fixed-amplitude common Goldstone modes of phases IIIA and IIIC, in which the same slowly varying internal rotation acts on all three Fourier fields.  Equal amplitudes alone do not ensure cancellation: generic phase-IIIB frame textures have label-dependent derivatives and remain anisotropic.  The anisotropy also survives in single- and double-$Q$ phases and near defect cores or walls.  Figure~\ref{fig:elasticity} summarizes this distinction geometrically.

Finally, consider terms bilinear in the order parameters and linear in one spatial derivative, conventionally called Lifshitz invariants.  Cross-label terms such as $\QQ_l\cdot\partial_i\QQ_m$ with $l\ne m$ are forbidden by translation symmetry.  The same-label scalar is a total derivative,
\begin{equation}
 \QQ_l\cdot\partial_i\QQ_l
 =\frac12\partial_i|\QQ_l|^2
 \label{eq:total_derivative}
\end{equation}
and therefore does not contribute to the bulk elastic energy.

For $N=2$, the remaining antisymmetric bilinear
\begin{equation}
 w_l^i=\epsilon_{ab}Q_l^a\partial_iQ_l^b
 \label{eq:planar_lifshitz}
\end{equation}
changes sign under an internal reflection and is forbidden by the independent $\OO(2)$ symmetry.  For $N=3$, $\QQ_l\times\partial_i\QQ_l$ is a spin-space axial vector, but the independent $\OO(3)$ symmetry supplies no invariant vector with which to contract it.  Thus no bulk Lifshitz invariant is allowed in the ideal spin-isotropic theory.  Dzyaloshinskii--Moriya terms can appear only when spin--orbit coupling reduces the independent spin symmetry and permits spin-space tensors tied to the lattice~\cite{Dzyaloshinsky1958,Moriya1960}.

When a Lifshitz contribution is present, its dynamical effect follows by embedding the continuum free energy in the Landau--Lifshitz equation,
\begin{equation}
 \partial_t\bm m=-\gamma\,\bm m\times\bm H_{\mathrm{eff}},
 \qquad
 \bm H_{\mathrm{eff}}\propto-\frac{\delta F}{\delta\bm m},
 \label{eq:landau_lifshitz}
\end{equation}
which converts the variational effective field into a chiral precessional torque~\cite{LandauLifshitz1935}.  Here $\bm m(\bm r,t)$ is the coarse-grained magnetization field reconstructed from the slowly varying Fourier amplitudes through Eq.~\ref{eq:spin_density}, $\gamma$ is its gyromagnetic ratio, and $F[\bm m]$ denotes the corresponding total magnetic free-energy functional.  In this work, the static functional is formulated in terms of the fields $\{\QQ_l\}$, and we do not analyze the resulting dynamics.

The elastic sector therefore affects defect profiles and energies without changing the homotopy classification.

\section{Discussion}
\label{sec:discussion}

\subsection{Free defects and wall-bound configurations}

Disconnected domain sectors are labeled by $\pi_0(\mathcal M)$, and domain walls interpolate between sectors represented by different elements of this set.  By contrast, a free vortex or $S^2$ texture is represented by a closed map into one connected component and requires no attached wall.  A circuit around a wall-bound endpoint instead maps to an open order-parameter path whose endpoints belong to different discrete-domain sectors and are joined only across the attached wall.  Table~\ref{tab:classification} contains three bulk defect classes: domain walls from disconnected components, vortices from nontrivial loops within one component, and $S^2$ textures.  Figure~\ref{fig:defect_atlas} identifies their phase-by-phase occurrence.  Wall-bound endpoints are distinct because they are not elements of the free bulk fundamental group.

\begin{figure}[tb]
\centering
\begin{tikzpicture}[x=1cm,y=1cm,font=\small,
  note/.append style={font=\small}]
  \coordinate (vh) at (-1.10,0);
  \node[font=\bfseries] at (0.15,1.65)
    {Wall-bound $\pi$ endpoint};
  \draw[qgray!60,dashed] (vh) circle (1.08);
  \filldraw[fill=white,draw=qgray,line width=0.8pt] (vh) circle (0.12);
  \foreach \a in {20,65,110,155,200,245,290,335}{
    \pgfmathsetmacro{\b}{\a/2}
    \draw[qarrow,qgreen]
      ($(vh)+(\a:0.72)$) -- ++(\b:0.34);
  }
  \draw[qorange,line width=2.0pt] (vh) -- ++(0:2.35);
  \draw[qorange!45,line width=0.45pt,dashed] (vh) -- ++(0:2.60);
  \node[note,text=qorange] at (0.35,0.30)
    {$\sigma_{\rm wall}>0$};
  \node[note,text=black] at (1.70,0.70) {domain $A$};
  \node[note,text=black] at (1.70,-0.58) {domain $B$};
  \node[note,text=qgreen] at (0.10,-1.45)
    {$\theta:0\rightarrow\pi$, closure requires $A\to B$};
\end{tikzpicture}
\caption{\label{fig:confined} A $\pi$ winding reaches a different antiphase or chirality component and becomes single valued only with a branch cut carrying a domain wall.  Unlike the free defects mapped in Table~\ref{tab:classification}, an isolated endpoint is linearly confined when the wall tension is nonzero.}
\end{figure}
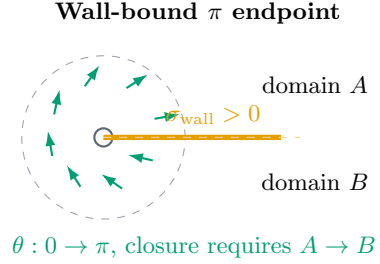

A fractional-looking winding can still occur locally, but it is not a free element of $\pi_1(\mathcal M_0)$.  A path that changes the spin angle by $\pi$ in phases IIIA$_{\OO(2)}$ or IIIC$_{\OO(2)}$ must be completed by a branch cut across which a discrete domain label changes.  For two wall endpoints with zero net $S^1$ winding separated by $L$, the asymptotic energy has the schematic form
\begin{equation}
 E(L)=2E_{\rm core}+A_q\rho_s\ln(L/a)
 +\sigma_{\rm wall}L+\mathcal O(1),
 \label{eq:confinement_energy}
\end{equation}
where $E_{\rm core}$ is the endpoint core energy, $a$ is a short-distance cutoff, $\rho_s$ is the relevant long-wavelength stiffness, and $A_q>0$ depends on the fractional winding and normalization.  The wall tension $\sigma_{\rm wall}>0$ produces the term linear in $L$, which dominates asymptotically and confines isolated endpoints.  Their detailed classification and stability belong to relative homotopy and defect energetics, not to the free bulk $\pi_1$ table.  Figure~\ref{fig:confined} shows this distinction schematically.

\subsection{Observable consequences}

Disconnected sectors can produce label-dependent signatures in reciprocal space.  An elementary $\Vfour$ wall reverses two Fourier components.  A distribution of finite antiphase domains can therefore reduce their phase coherence and produce label-dependent peak broadening or diffuse scattering, without implying a universal change in integrated intensity.

The label-locked stiffness can produce direction-dependent texture widths or core shapes in single- and double-$Q$ phases and in generic phase-IIIB frame textures.  This anisotropy cancels from the fixed-amplitude far-field common modes of phases IIIA and IIIC.  Chirality-sensitive probes are phase dependent: the IIIB scalar chirality can couple to Hall responses in conducting systems~\cite{Park2023}, whereas the IIIC vector chirality is naturally probed by polarization-sensitive magnetic scattering~\cite{JinExperiment2025,Blume1963}.

Finally, real-space imaging of an apparent $\pi$ winding should reveal a terminating discrete-domain wall or a second endpoint connected by such a wall.  This attachment rule follows directly from the topology; quantitative line shapes, texture widths, and core profiles remain nonuniversal.

\subsection{Role of spin--orbit coupling}

The ideal theory assumes the direct-product symmetry $S\times\OO(N)$, with spin rotations independent of the lattice.  Spin--orbit coupling breaks this independence by tying spin transformations to crystalline operations.  If the resulting spin--orbit-induced terms are weak, Table~\ref{tab:classification} can still organize defect cores and intermediate-scale textures on length scales shorter than the spin-pinning length.  At asymptotically large distances, however, even weak anisotropy changes the vacuum manifold and hence the exact defect classification.

When spin--lattice locking must be treated explicitly, the parent symmetry is the actual magnetic symmetry group rather than $S\times\OO(N)$.  Its identity component $G_0$, stabilizer $H$, and connected stabilizer $K=H\cap G_0$ must all be recomputed.  The spin--orbit-coupled classification therefore cannot be obtained merely by adding an anisotropy energy while retaining the manifolds in Table~\ref{tab:classification}.

\subsection{Dimensional and multi-sublattice scope}

Because the classification is determined by the internal ordered-state manifold, the same manifolds apply in two and three spatial dimensions whenever the corresponding ordered state is well defined.  Their physical interpretation depends on dimension: $\pi_0$ gives line walls in two dimensions and surface walls in three, $\pi_1$ gives point vortices in two and vortex lines in three, and $\pi_2$ gives skyrmion textures in two and hedgehog defects in three.  In a strictly two-dimensional system with exact continuous spin symmetry, this statement concerns the zero-temperature ordered manifold or finite length scales over which the order parameter remains well defined; finite-temperature ordering requires the usual qualifications.  The elastic functional derived in Sec.~\ref{sec:elasticity} is explicitly two dimensional; a stacked three-dimensional system requires additional gradients along the stacking direction.

The table applies directly to the one-mode-per-$M_l$ theory defined in Sec.~\ref{subsec:convention}.  In a honeycomb, kagome, or other multi-sublattice realization, retaining additional sublattice form factors or irreducible-representation modes enlarges $\Phi$ beyond three $N$-component vectors.  Symmetry-allowed couplings among these modes can modify the minimizing configurations and the stabilizers $H$ and $K$, so the classification must then be recomputed in the enlarged order-parameter space.

\subsection{How non-Abelian charges can reappear}

Phase IIIB has $\mathcal M_0\simeq\SO(3)$ and therefore the Abelian fundamental group $\pi_1(\mathcal M_0)=\ZZ_2$.  Non-Abelian vortex composition requires a different physical configuration space with a noncommutative fundamental group.

One possibility is a genuinely unlabeled frame with a physical discrete identification $K_{\mathrm{frame}}\subset\SO(3)$.  Its manifold is $\SO(3)/K_{\mathrm{frame}}$, and its fundamental group is the preimage of $K_{\mathrm{frame}}$ under the covering $\SU(2)\to\SO(3)$; this group is non-Abelian for identifications such as the tetrahedral group.  A second possibility is to admit lattice dislocations or disclinations, allowing magnetic paths to close jointly with lattice holonomy.  These are combined magnetic--crystalline defects rather than free internal vortices and are not automatically non-Abelian.  Finally, spin--lattice locking changes the magnetic parent group and therefore changes $G_0$, $H$, and $K=H\cap G_0$.  Its effect on vortex composition is model dependent: $\pi_1(G_0/K)$ must be computed case by case and is non-Abelian only when this fundamental group is noncommutative.

None of these possibilities follows from reinterpreting the fixed-lattice quotient used in Table~\ref{tab:classification}.

\section{Conclusion}
\label{sec:conclusion}

We have classified the topological defects and textures governed by $\pi_0$, $\pi_1$, and $\pi_2$ in all seven stable phases of the ideal $(\Vfour\rtimes\Dthree)\times\OO(N)$ fixed-lattice theory.  The full stabilizer $H$ and $K=H\cap G_0$ are essential because combined crystalline--spin operations relate distinct domains without identifying configurations within a connected component.  The manifolds support $Q$-selection, antiphase, chirality, and frame-orientation walls; integer planar and Abelian $\ZZ_2$ frame vortices; and $S^2$ textures.  Thus apparent half windings are wall-bound, and phase IIIB has an Abelian rather than a non-Abelian binary-polyhedral vortex group.  Translation symmetry removes cross-label quadratic gradients, point-group symmetry leaves two same-label stiffnesses, and independent $\OO(N)$ symmetry excludes bulk Lifshitz invariants.  Extensions include spin--orbit locking, core energetics, and scattering signatures; wall-bound $\pi$ endpoints and Abelian phase-IIIB vortex composition provide direct experimental tests.

\begin{acknowledgments}
This work was supported in part by the National Key Research and Development Program of China (Grant No. 2022YFA1403403) and the National Natural Science Foundation of China (Grant Nos. 12274441 and 12534004). J.-T.J is supported by a fellowship award from Hong Kong Research Grant Council (Project No. SRFS2324-6S01) and New Conerstone Science Foundation.
\end{acknowledgments}

\section*{Data Availability}
No data were created or analyzed in this study.

\appendix

\section{Graph stabilizer of phase IIB}
\label{app:IIB}

For the reference state
\begin{equation}
 \Phi_0=Q_0(\hat{\bm e}_1\ \hat{\bm e}_2\ \bm0),
 \label{eq:IIB_appendix_reference}
\end{equation}
the $M_3$ amplitude vanishes, while the occupied $M_1$ and $M_2$ columns form an ordered orthonormal frame.  The internal transformations that map this frame to a signed permutation of itself form the eight-element group
\begin{equation}
 \mathcal P_2=
 \left\{
 \begin{pmatrix}\epsilon_1&0\\0&\epsilon_2\end{pmatrix},
 \begin{pmatrix}0&\epsilon_1\\\epsilon_2&0\end{pmatrix}
 \mathrel{\Big|}\epsilon_1,\epsilon_2=\pm1
 \right\}\subset\OO(2).
 \label{eq:P2}
\end{equation}
The crystalline subgroup
\[
 S_{M_3}=\{s\in S\mid s\text{ preserves the }M_3\text{ label}\}
\]
also has eight elements.  Its signed action on the occupied $M_1$ and $M_2$ columns gives a bijection from $S_{M_3}$ to $\mathcal P_2$.  Hence, for each $U\in\mathcal P_2$, there is a unique $s_U\in S_{M_3}$ satisfying
\begin{equation}
 \Phi_0\rho(s_U)=U\Phi_0.
 \label{eq:IIB_compensation}
\end{equation}
The full stabilizer is therefore the graph of this bijection,
\begin{equation}
 H_{\mathrm{IIB}}=\{(s_U,U)\mid U\in\mathcal P_2\}.
 \label{eq:IIB_graph}
\end{equation}
Only $U=I$ is paired with $s_U=e$.  Hence the connected stabilizer is
\[
 K=H_{\mathrm{IIB}}\cap G_0=\{(e,I)\}.
\]
Since $S_{M_3}$ has index three in $S$, the crystalline orbit contains three choices of the missing $M$ point.  For each choice, the $\OO(2)$ spin orbit has two frame-orientation components, each homeomorphic to $S^1$.  Therefore
\begin{equation}
 \mathcal M_{\mathrm{IIB},\OO(2)}
 =\bigsqcup_3\OO(2)=\bigsqcup_6S^1,
 \label{eq:IIB_appendix_manifold}
\end{equation}
as quoted in Eq.~\eqref{eq:IIB_manifold}.

\section{Elastic invariance equations}
\label{app:gradient}

Define
\[
 t_{l,ij}=(\partial_i\QQ_l)\cdot(\partial_j\QQ_l).
\]
The same-label quadratic elastic energy can then be parameterized as
\begin{align}
 f_{\nabla}^{(2)}
 &=\sum_{l=1}^3
 \left(a_lt_{l,xx}+b_lt_{l,xy}+c_lt_{l,yy}\right),
 \label{eq:gradient_coefficients}\\
 \bm k_l&=(a_l,b_l,c_l)^T.
 \nonumber
\end{align}
Here $\bm k_l$ collects the three spatial coefficients associated with label $l$.
Writing
\[
 \bm t_l=(t_{l,xx},t_{l,xy},t_{l,yy})^T,
\]
the reflection $A$ in Eq.~\eqref{eq:gradient_A} acts within each label as
\begin{equation}
 \bm t_l\mapsto C_A\bm t_l,\qquad
 C_A=\operatorname{diag}(1,-1,1),
 \label{eq:CA}
\end{equation}
and simultaneously exchanges labels $2$ and $3$.  Since
$f_{\nabla}^{(2)}=\sum_l\bm k_l^T\bm t_l$, invariance requires
\begin{equation}
 C_A^T\bm k_1=\bm k_1,\qquad
 C_A^T\bm k_2=\bm k_3,\qquad
 C_A^T\bm k_3=\bm k_2.
 \label{eq:A_invariance}
\end{equation}
Under the spatial derivative transformation associated with the
$120^\circ$ operation $D$, each $\bm t_l$ is mapped to $B\bm t_l$, where
\begin{equation}
 B=
 \begin{pmatrix}
  1/4&\sqrt3/2&3/4\\
  -\sqrt3/4&-1/2&\sqrt3/4\\
  3/4&-\sqrt3/2&1/4
 \end{pmatrix}.
 \label{eq:Bmatrix}
\end{equation}
Including the cyclic permutation of the Fourier labels, the complete action is
\[
 (\bm t_1,\bm t_2,\bm t_3)
 \mapsto(B\bm t_2,B\bm t_3,B\bm t_1).
\]
The transformed energy is therefore
\[
 \bm k_1^TB\bm t_2+\bm k_2^TB\bm t_3+\bm k_3^TB\bm t_1,
\]
so invariance requires
\begin{equation}
 B^T\bm k_1=\bm k_2,\qquad
 B^T\bm k_2=\bm k_3,\qquad
 B^T\bm k_3=\bm k_1.
 \label{eq:D_invariance}
\end{equation}
Collecting the reflection and rotation constraints in the ordered coefficient
vector
\[
 \bm c=(a_1,a_2,a_3,b_1,b_2,b_3,c_1,c_2,c_3)^T
\]
reproduces the aggregate equations
$(\mathsf A-I)\bm c=(\mathsf D-I)\bm c=0$ in
Eq.~\eqref{eq:gradient_nullspace}.  Row reduction gives a
two-dimensional solution space spanned by
\begin{align}
 \bm v_1&=(1,1,1,0,0,0,1,1,1)^T,\nonumber\\
 \bm v_2&=(2,-1,-1,0,2\sqrt3,-2\sqrt3,-2,1,1)^T.
 \label{eq:gradient_kernel_vectors}
\end{align}
With the bilinear ordering in Eq.~\eqref{eq:gradient_basis},
$\bm v_1\cdot\bm t=\mathcal I_1$ and
$\bm v_2\cdot\bm t=2\mathcal I_2$.  Thus the most general invariant is
a linear combination of Eqs.~\eqref{eq:I1} and~\eqref{eq:I2}.

\bibliographystyle{apsrev4-2}
\bibliography{triple-Q}

\end{document}